\documentclass[11pt]{article}
\usepackage[a4paper]{geometry}
\usepackage{amsfonts, amsmath, amssymb, amsthm, graphicx, caption, authblk, multirow, makecell, framed, float, xcolor, enumitem, tikz, hyperref}
\usepackage{algorithm,algorithmicx,algpseudocode}
\usetikzlibrary{fit,shapes.arrows}
\theoremstyle{definition}

\theoremstyle{remark}

\definecolor{blk}{RGB}{63,63,63}
\newcommand*{\mybox}[1]{%
  \framebox{\raisebox{0cm}[0.5\baselineskip][0.05\baselineskip]{%
    \hbox to 0.10cm {\hss#1\hss}}}\hspace{0.05cm}}

\begin{document}
\title{Balance-Based Cryptography: Physically Computing Any Boolean Function}
\author[1]{Suthee Ruangwises\thanks{\texttt{suthee@cp.eng.chula.ac.th}}}
\affil[1]{Department of Computer Engineering, Faculty of Engineering, Chulalongkorn University, Bangkok, Thailand}
\date{}
\maketitle

\begin{abstract}
Secure multi-party computation is an area in cryptography which studies how multiple parties can compare their private information without revealing it. Besides digital protocols, many unconventional protocols for secure multi-party computation using physical objects have also been developed. The vast majority of them use playing cards as the main tools. In 2024, Kaneko et al. introduced the use of a balance scale and coins in zero-knowledge proof protocols for pencil puzzles. In this paper, we extend the use of these tools to secure multi-party computation. In particular, we develop four protocols that can securely compute any $n$-variable Boolean function using a balance scale and coins.

\end{abstract}

\section{Introduction}
Secure multi-party computation (MPC) is one of the most actively studied areas in cryptography. It investigates how multiple parties can compare their private information without revealing it.

One of the best-known classical examples of MPC involves the following problem. Alice and Bob want to know whether they both like each other. However, no one wants to confess first due to fear of embarrassment of getting rejected. They need a protocol that only distinguishes the two cases where they both like each other and otherwise without leaking any other information. Theoretically, this setting is equivalent to computing a logical AND function of two input bits, one from each player.

Besides the AND function, other widely studied Boolean functions include a logical XOR function, a \textit{majority function} (deciding whether there are more 1s than 0s in the inputs), and an \textit{equality function} (deciding whether all inputs are equal).

Instead of digital protocols, many researchers have developed unconventional protocols for MPC using physical objects found in everyday life such as cards, coins, and envelopes. These protocols have the benefit that they do not require computers and also allow external observers to verify that all parties truthfully execute them (which is often a challenging task for digital protocols). In addition, they are easier to understand and to verify the correctness and security, even for non-experts, and thus can be used for didactic purposes.

The vast majority of existing physical protocols use playing cards as the main tools, thus this area of research is often called \textit{card-based cryptography}. During the past decade, card-based cryptography has rapidly gained interest and has been extensively studied, with dozens of papers published \cite{landscape2,landscape1,landscape3}.

\subsection{Related Work}
Research in card-based cryptography began in 1989 when den Boer \cite{5card} introduced the \textit{five-card trick} protocol to solve the classical Alice-and-Bob problem. This protocol can compute the AND function of two input bits from two players using five cards: three identical black cards and two identical red cards.

Mizuki and Sone \cite{mizuki09} later improved the AND protocol to accommodate $n$ inputs. Apart from the AND function, card-based protocols to compute other Boolean functions have also been proposed, including the XOR function \cite{mizuki09}, the majority function \cite{nishida13,toyoda}, and the equality function \cite{ruangwises21}. Nishida et al. \cite{nishida15} proved that any $n$-variable Boolean function can be computed using $2n+6$ cards, and any such function that is symmetric can be computed using $2n+2$ cards.

Besides cards, other physical objects such as coins \cite{coin,coin2}, envelopes \cite{envelope}, combination locks \cite{diallock}, balls and bags \cite{balls}, printed transparencies \cite{printed}, and PEZ candy dispensers \cite{pez2,pez1,pez3} have also been used in cryptographic protocols.

In 2024, Kaneko et al. \cite{balance} were the first ones to use a balance scale and coins in a cryptographic protocol.\footnote{While some earlier protocols \cite{coin,coin2} also used coins, only faces (head or tail) were used to encode inputs without considering the weight of the coins.} Their protocols, however, were not designed to compute functions in the MPC setting, but are zero-knowledge proof protocols to verify solutions of pencil puzzles like Sudoku. Very recently, Kaneko et al.~\cite{balance2} also proposed a voting protocol using the same tools, which can compute a specific function in the MPC setting.

\subsection{Our Contribution}
In this paper, we extend the use of a balance scale and coins as physical tools to securely compute arbitrary Boolean functions, providing a new physical model for a generic MPC setting. Note that coins can be replaced by any small objects with significant weights like balls or marbles, but we use coins to preserve the setting proposed by Kaneko et al. \cite{balance}.

In particular, we develop four protocols that securely compute $n$-variable Boolean functions using a balance scale and coins. The first protocol can compute the AND function. The second one can compute any \textit{threshold function} (including the AND function and the majority function). The third protocol can compute any symmetric function, while the fourth one can compute any Boolean function. See Table \ref{table0} for the number of coins, bags, and comparisons (times using the balance scale to weigh objects) required for each protocol.

Our protocols are also practical. Unlike the two existing balance-based protocols \cite{balance,balance2} which use same-size coins with $n$ different weights, our protocols only use coins with two different weights, removing a practical challenge to implement them in real world.

\begin{table}
	\centering
	\caption{Number of coins, bags, and comparisons required for each protocol to compute an $n$-variable Boolean function} \label{table0}
	\begin{tabular}{|c|c|c|c|c|c|}
		\hline
		\textbf{Protocol} & \textbf{Function} & \textbf{\#Coins} & \textbf{\#Bags} & \textbf{\#Comparisons} & \textbf{Notes} \\ \hline
		Protocol 1 (\S\ref{prot1}) & AND & $3n$ & 0 & 1 & \\ \hline
		Protocol 2 (\S\ref{prot2}) & threshold & $2n$ & 0 & 1 & Uses a custom weight. \\ \hline
		Protocol 3 (\S\ref{prot3}) & symmetric & \thead{at most\\ $n\lceil \frac{n}{2}+2 \rceil$} & \thead{at most\\ $\lceil \frac{n}{2}+1 \rceil$} & at most $\lceil \frac{n}{2} \rceil$ & Uses a pen. \\ \hline
		Protocol 4 (\S\ref{prot4}) & any & \thead{at most\\ $n(2^n+1)$} & \thead{at most\\ $2^{n-1}$} & at most $2^{n-1}$ & \\ \hline
	\end{tabular}
\end{table}

\section{Preliminaries}
\subsection{Balance Scale}
Our protocols use a double-pan balance scale (see Fig. \ref{fig1}). When placing objects on both pans, the scale will tilt towards the heavier side at a constant angle (regardless of weight difference), or will stay put if the objects on both pans have equal weight.

Note that in reality, each balance scale may have a tolerance $\varepsilon$, where the scale will stay put (due to friction) if the difference of weights on both pans is at most $\varepsilon$.

\subsection{Coins}
\begin{figure}
\captionsetup{width=0.8\textwidth}
\centering
\includegraphics[width=50mm]{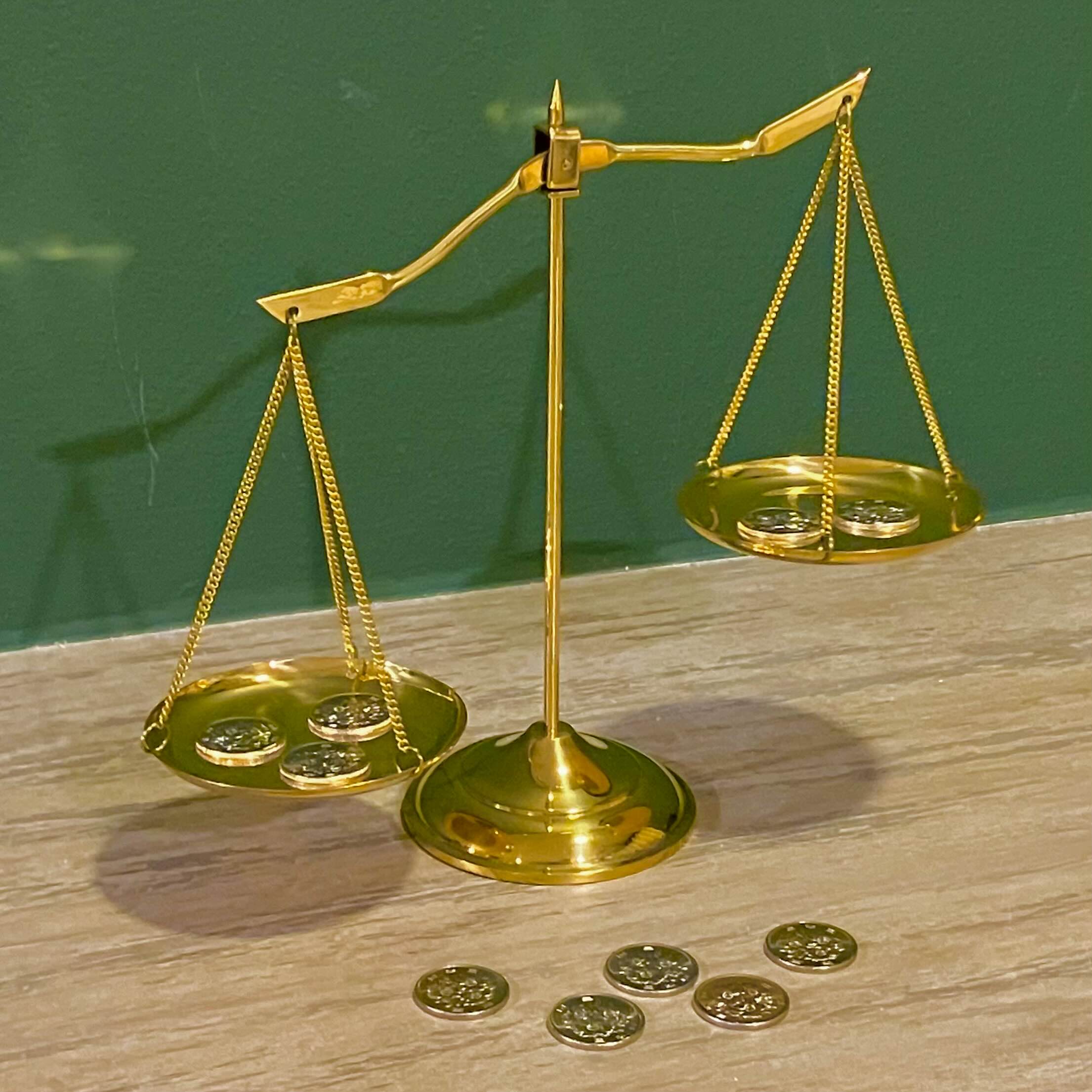} \hspace{2mm}
\includegraphics[width=50mm]{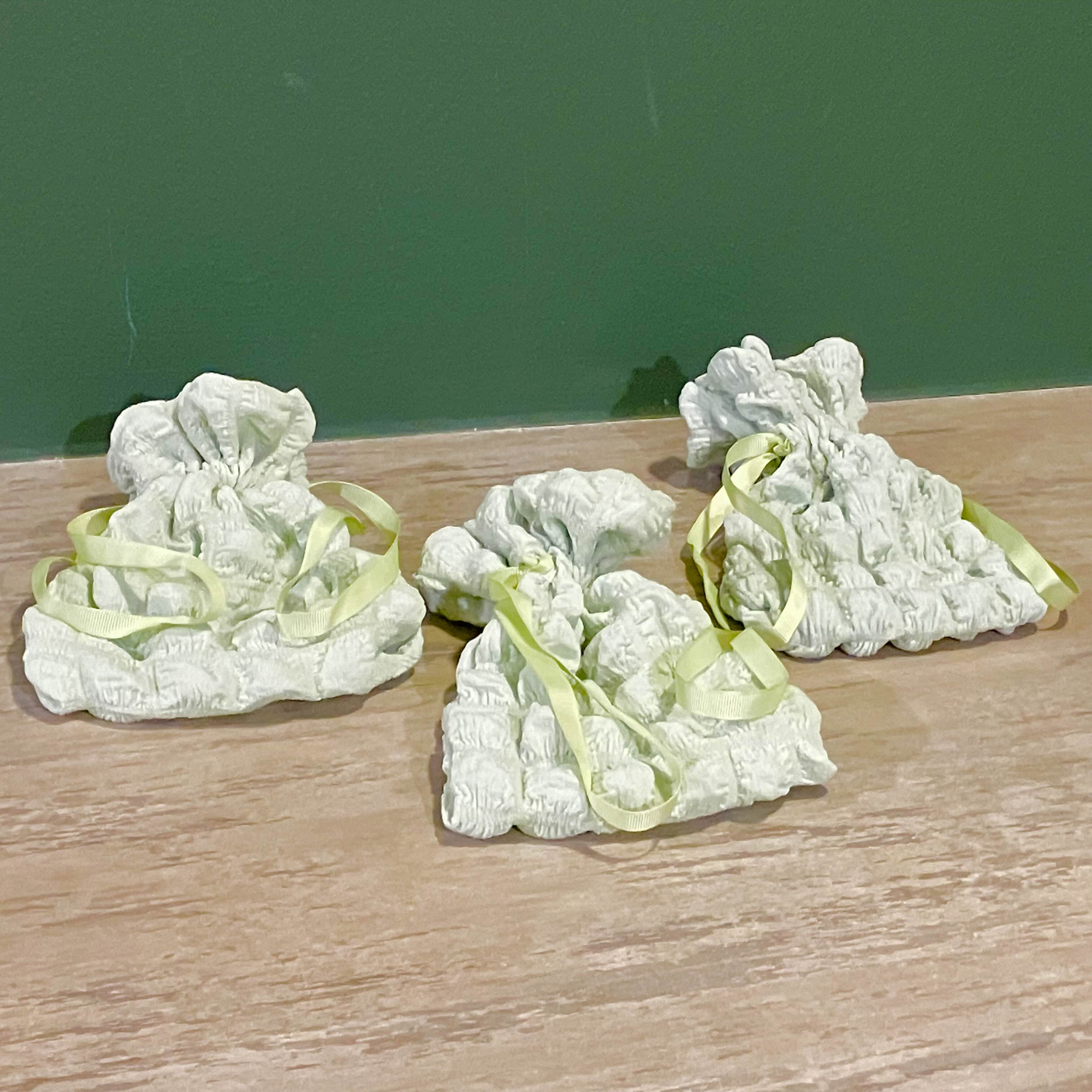}
\caption{Examples of a balance scale and coins (left) and bags (right) used in our protocols} \label{fig1}
\end{figure}

Two types of coin, called \textit{heavy coins} and \textit{light coins}, are used in our protocols. All coins have the same size and color (see Fig. \ref{fig1}). Each heavy coin weighs $w$, while each light coin weighs $w-\delta$. We assume that $\delta$ is much larger than $\varepsilon$, so the balance scale can detect the difference between heavy and light coins.

\subsection{Custom Weights}
A custom weight is only used in the threshold function protocol in Section \ref{prot2}. The exact weight depends on the function being computed, but is always in the form $pw+q(w-\delta)+\frac{\delta}{2}$, where $p,q \in \mathbb{Z}^+ \cup \{0\}$.

When computing multiple different threshold functions, instead of preparing a lot of different custom weights, it is sufficient to prepare only one custom weight of $\frac{\delta}{2}$, as all weights in such form can be obtained by combining this custom weight with $p$ heavy coins and $q$ light coins.

\subsection{Bags and Shuffle Operation}
Bags used in our protocols are small opaque bags with the same size, color, and weight, and can be closed at the top (see Fig. \ref{fig1}). Each bag can contain multiple coins and can be placed on a balance scale.

Given $k$ bags, each with the same number of coins inside (but with slightly different weights), the players may jointly \textit{shuffle} all bags into a uniformly random permutation unknown to all players. This operation has previously been used in \cite{balls}, where there are balls inside each bag.

The shuffle operation can be done in real world by all players scrambling the bags together on a table. We assume that no player can determine the exact weight of any bag just by touching it.

\section{Computing AND Function} \label{prot1}
First, we propose the following protocol for one of the most fundamental functions: the AND function. In a setting with $n$ players, where the $i$-th player has a bit $x_i$, this protocol securely computes $x_1 \wedge x_2 \wedge ... \wedge x_n$.

Each player is given a heavy coin (with weight $w$) and a light coin (with weight $w-\delta$). Each player then chooses one of the two coins to place on the same side of a balance scale. If the player's bit is 1, the player chooses the heavy coin; if the player's bit is 0, the player chooses the light coin.

We now have $n$ coins placed on one side of the scale. Then, the players jointly place $n$ heavy coins on the other side of the scale. If the weights of the two sides are equal, return 1; otherwise, return 0.

This protocol is correct because the only case where the weights are equal is when all players' coins are heavy coins (and the players' coins are lighter in all other cases), which occurs when all players' bits are 1s. Moreover, each player learns no information other than whether the $n$ coins combined have weight $nw$ or less than $nw$. Therefore, the protocol is correct and secure. This protocol uses $3n$ coins (two given to each player plus $n$ additional heavy coins) and uses one comparison.

\section{Computing Threshold Function} \label{prot2}
For a positive integer $k \leq n$ and bits $x_1,x_2,...,x_n \in \{0,1\}$, a Boolean \textit{threshold function} $T_k$ returns 1 if and only if at least $k$ of the $n$ input bits are 1s. Formally,

$$T_k(x_1,x_2,...,x_n) :=
\begin{cases}
	1, &\text{if $\sum_{i=1}^n x_i \geq k$;} \\
	0, &\text{otherwise.}
\end{cases}$$

Note that $T_k$ is the AND function when $k=n$, and is the majority function when $k=\lfloor \frac{n}{2} \rfloor + 1$.

We can slightly modify the AND protocol in Section \ref{prot1} to compute any threshold function. Similarly to the AND protocol, each player is given a heavy coin and a light coin, and then chooses the heavy (resp. light) coin if the player's bit is 1 (resp. 0) to place on the same side of a balance scale.

The players then place a custom weight $W = (k-1)w+(n-k+1)(w-\delta)+\frac{\delta}{2}$ on the other side of the scale.\footnote{Alternatively, the weight $W$ can be obtained by placing $k-1$ heavy coins, $n-k+1$ light coins, and a custom weight of $\frac{\delta}{2}$.} If the $n$ coins combined are heavier than $W$, return 1; otherwise, return 0.

Observe that $W$ is heavier than $k-1$ heavy coins plus $n-k+1$ light coins, but is lighter than $k$ heavy coins plus $n-k$ light coins. As a result, the $n$ coins are heavier than $W$ if and only if at least $k$ of them are heavy coins (and the $n$ coins are lighter in all other cases). This occurs when at least $k$ of the players' bits are 1s, i.e. $\sum_{i=1}^n x_i \geq k$, which is the exact condition for $T_k(x_1,x_2,...,x_n)=1$. Moreover, each player learns no information other than whether the $n$ coins are heavier or lighter than $W$. Therefore, the protocol is correct and secure.

This protocol uses $2n$ coins and a custom weight $W$ (or $3n$ coins and a custom weight of $\frac{\delta}{2}$), and uses one comparison.

\section{Computing Symmetric Function} \label{prot3}
A Boolean function $f:\{0,1\}^n \rightarrow \{0,1\}$ is called symmetric if $$f(x_1,x_2,...,x_n)=f(x_{\sigma(1)},x_{\sigma(2)},...,x_{\sigma(n)})$$ for any $x_1,x_2,...,x_n \in \{0,1\}$ and any permutation $\sigma: \{1,2,...,n\} \rightarrow \{1,2,...,n\}$. Note that for any such function $f$, the value of $f(x_1,x_2,...,x_n)$ only depends on the number of inputs being 1s, i.e. the integer sum $\sum_{i=1}^n x_i$.

We denote an $n$-variable symmetric Boolean function by $S_X^n$ for some $X \subseteq \{0,1,...,n\}$. A function $S_X^n$ is defined by

$$S_X^n(x_1,x_2,...,x_n) :=
\begin{cases}
	1, &\text{if $\sum_{i=1}^n x_i \in X$;} \\
  0, &\text{otherwise.}
\end{cases}$$

For example, the AND function $x_1 \wedge x_2 \wedge ... \wedge x_n$ is denoted by $S_{\{n\}}^n$, and the XOR function $x_1 \oplus x_2 \oplus ... \oplus x_n$ is denoted by $S_{\{1,3,5,...,2\lceil\frac{n}{2}\rceil-1\}}^n$.

To compute $S_X^n$, we prepare $|X|$ bags, each one corresponding to each element of $X$. A bag corresponding to $k \in X$ contains $k$ heavy coins and $n-k$ light coins.

We also prepare an additional bag, called a \textit{special bag}, for players to put coins in. Each player is given a heavy coin and a light coin, and then chooses the heavy (resp. light) coin if the player's bit is 1 (resp. 0) to put in this bag.

The players use a pen to make a small mark inside of the special bag. This mark must not visible from the outside without opening the bag. After that, all players together shuffle all $|X|$ non-special bags randomly. Then, the players jointly perform the following steps.

\begin{enumerate}
	\item Pick an arbitrary non-special bag. Shuffle this bag with the special bag randomly.
	\item Place each of these two bags on each side of the scale. If the weights of the two sides are equal, return 1 and end the protocol.
	\item Take both bags from the scale. Shuffle them randomly again.
	\item Open both bags and find a pen mark to locate the special bag. Then, close both bags and disregard the non-special bag.
\end{enumerate}

Repeat the above steps for a new non-special bag. After $|X|$ iterations, if no non-special bag has weight equal to the special bag, return 0.

\subsection{Proofs of Correctness and Security}
The protocol returns 1 when there is a non-special bag with weight equal to the special bag (and there can be at most one such bag). Suppose this non-special bag corresponds to an element $k \in X$ (containing $k$ heavy coins and $n-k$ light coins). This case occurs if and only if exactly $k$ of the $n$ players' coins are heavy coins, which is equivalent to $\sum_{i=1}^n x_i = k$. Therefore, the protocol returns 1 if and only if $\sum_{i=1}^n x_i \in X$, which is the exact condition for $S_X^n(x_1,x_2,...,x_n) = 1$.

Consider the process of comparing a special bag and a non-special bag in each iteration. Due to the shuffles both before and after weighing, no player can distinguish which side of the scale is a special bag. Therefore, all information each player learns is whether or not both bags have equal weight.

Also, due to the shuffle of all $|X|$ non-special bags into a uniformly random permutation, no player can learn about which non-special bag corresponds to which element of $X$. Therefore, all information each player learns is whether or not there is a non-special bag with weight equal to the special bag. Therefore, the protocol is correct and secure.

\subsection{Optimization}
In the protocol, we need $|X|+1$ bags to compute $S_X^n$. However, if $|X| > \frac{n+1}{2}$, we can instead compute the function $S_{\{0,1,...,n\}-X}^n$ and then negate the output. In such cases, we have $|\{0,1,...,n\}-X| = n+1-|X| < \frac{n+1}{2}$. Therefore, the number of required bags can be reduced to at most $\lfloor \frac{n+1}{2} \rfloor + 1 = \lceil \frac{n}{2} \rceil + 1$.

The number of comparisons is equal to the number of non-special bags $|X|$, which can be reduced to at most $\lceil \frac{n}{2} \rceil$. The number of required coins is $n(|X|+2)$ (two given to each player plus $n$ in each non-special bag), which can be reduced to at most $n(\lceil \frac{n}{2} \rceil + 2)$.

\section{Computing Any Boolean Function} \label{prot4}
Finally, we propose the following protocol to compute an arbitrary $n$-variable Boolean function among $n$ players, where the $i$-th player has a bit $x_i$.

For any Boolean function $f: \{0,1\}^n \rightarrow \{0,1\}$, we can write $f$ as Boole's expansion (or Shannon expansion) \cite{boole}:

\begin{align*}
f(x_1,x_2,...,x_n) &= \bar{x}_1\bar{x}_2...\bar{x}_nf(0,0,...,0) \vee x_1\bar{x}_2...\bar{x}_nf(1,0,...,0) \\
&\vee \bar{x}_1x_2...\bar{x}_nf(0,1,...,0) \vee x_1x_2...\bar{x}_nf(1,1,...,0) \\
&\vee ... \vee x_1x_2...x_nf(1,1,...,1) \\
&= \bigvee_{(b_1,b_2,...,b_n) \in \{0,1\}^n} (x_1 \leftrightarrow b_1)(x_2 \leftrightarrow b_2)...(x_n \leftrightarrow b_n)f(b_1,b_2,...,b_n),
\end{align*}

where $\vee$ and $\leftrightarrow$ are Boolean OR and XNOR (i.e. if and only if) operations, respectively. Observe that we only need to consider the terms where $f(b_1,b_2,...,b_n)=1$, since the rest of the terms are all zeroes. The output $f(x_1,x_2,...,x_n)$ is 1 if and only if at least one such term is 1.

Let $B_f=\{(b_1,b_2,...,b_n) \in \{0,1\}^n | f(b_1,b_2,...,b_n)=1\}$. We prepare $|B_f|$ bags, each one corresponding to each element of $B_f$. Also, each player is given $2^{n-1}$ heavy coins and $2^{n-1}$ light coins.

For each bag corresponding to $(b_1,b_2,...,b_n)$, the $i$-th player puts into that bag a heavy coin if $x_i=b_i$, and a light coin if $x_i \neq b_i$.\footnote{We assume a \textit{semi-honest model} \cite{practical}, where each player puts a coin into every bag consistently according to the rule and does not try to sabotage or hack the protocol.}

After that, all players together shuffle all $|B_f|$ bags randomly. Then, for each bag, they pour the coins inside it on one side of a balance scale, and put $n$ heavy coins on the other side. Repeat this for every bag. If there is a bag where the weights of the two sides are equal, return 1; otherwise, return 0.

Note that for a special case where $|B_f|=1$, a bag is not necessary; the players can directly place coins on the scale similarly to the AND protocol.

\subsection{Example: Three-Input XOR Function} \label{ex1}
To clearly illustrate the protocol, we show an example to compute the three-variable XOR function $f(x_1,x_2,x_3) = x_1 \oplus x_2 \oplus x_3$.

First, prepare four bags corresponding to (1,0,0), (0,1,0), (0,0,1), and (1,1,1), respectively. Each player is given two heavy coins and two light coins.

Each player then puts a coin in each bag according to Table \ref{table1}. For instance, if $x_1=1$, then Player 1 puts a heavy coin in Bag (1,0,0) and Bag (1,1,1), and a light coin in Bag (0,1,0) and Bag (0,0,1).

\begin{table}[H]
	\centering
	\caption{The coin each player putting in each bag} \label{table1}
	\begin{tabular}{|c|c|c|c|c|c|}
		\hline
		\multicolumn{2}{|c|}{\multirow{2}{*}{\textbf{Player}}} & \multicolumn{4}{c|}{\textbf{Bag}} \\ \cline{3-6}
		\multicolumn{2}{|c|}{} & (1,0,0) & (0,1,0) & (0,0,1) & (1,1,1) \\ \hline \hline
		\multirow{2}{*}{\textbf{Player 1}} & $x_1=1$ & heavy & light & light & heavy \\ \cline{2-6}
		& $x_1=0$ & light & heavy & heavy & light \\ \hline \hline
		\multirow{2}{*}{\textbf{Player 2}} & $x_2=1$ & light & heavy & light & heavy \\ \cline{2-6}
		& $x_2=0$ & heavy & light & heavy & light \\ \hline \hline
		\multirow{2}{*}{\textbf{Player 3}} & $x_3=1$ & light & light & heavy & heavy \\ \cline{2-6}
		& $x_3=0$ & heavy & heavy & light & light \\ \hline
	\end{tabular}
\end{table}

After that, all players together shuffle all four bags randomly. Then, for each bag, they pour the coins inside it on one side of a balance scale, and put three heavy coins on the other side. Repeat this for every bag. If there is a bag where the weights of the two sides are equal, return 1; otherwise, return 0.

\subsection{Proofs of Correctness and Security}
The protocol returns 1 when there is a bag corresponding to $(b_1,b_2,...,b_n) \in B_f$ containing $n$ heavy coins, which occurs when $x_i=b_i$ for every $i=1,2,...,n$ (and there can be at most one such bag). Therefore, the protocol returns 1 if and only if $(x_1,x_2,...,x_n) \in B_f$, which is the exact condition for $f(x_1,x_2,...,x_n)=1$.

Consider the process of comparing $n$ heavy coins with the $n$ coins inside each bag. Each player learns no information other than whether the $n$ coins inside the bag have weight $nw$ or less than $nw$, i.e. whether or not they are all heavy coins. Also, due to the shuffle of all $|B_f|$ bags into a uniformly random permutation, no player can learn about which bag correspond to which element of $B_f$. Therefore, all information each player learns is whether or not there is a bag containing all heavy coins. Therefore, the protocol is correct and secure.

\subsection{Optimization}
In the protocol, we need $|B_f|$ bags to compute $f$. However, if $|B_f| > 2^{n-1}$, we can instead compute the function $\bar{f}(x_1,x_2,...,x_n) := 1-f(x_1,x_2,...,x_n)$ and then negate the output. In such cases, we have $|B_{\bar{f}}| = 2^n-|B_f| < 2^{n-1}$. Therefore, the number of required bags can be reduced to at most $2^{n-1}$. As the number of comparisons is equal to the number of bags, it can also be reduced to at most $2^{n-1}$.

For $1 \leq i \leq n$, let $p_i$ (resp. $q_i$) be the numbers of $(b_1,b_2,...,b_n) \in B_f$ such that $b_i=0$ (resp. $b_i=1$). Observe that the $i$-th player has to use at most $\max(p_i,q_i)$ heavy coins and at most $\max(p_i,q_i)$ light coins (we have $\max(p_i,q_i) \leq |B_f| \leq 2^{n-1}$). As a result, we can give only $\max(p_i,q_i)$ heavy coins and $\max(p_i,q_i)$ light coins to the $i$-th player (instead of the maximum $2^{n-1}$ heavy coins and $2^{n-1}$ light coins). For instance, in the example in Section \ref{ex1}, each player receives only two heavy coins and two light coins (instead of four heavy coins and four light coins). The total number of required coins is at most $n(2^n+1)$ (at most $2^n$ given to each player plus $n$ additional heavy coins).

\section{Conclusion}
We developed four balance-based protocols that can compute arbitrary $n$-variable Boolean functions, and proved the correctness and security of these protocols. By doing so, we extended the use of a balance scale and coins to a generic MPC setting, implying that they have computational power equal to cards, PEZ candy dispensers, and other existing physical tools.

\end{document}